\documentclass[pra,aps,twocolumn,amsmath,amssymb,floatfix,reprint,footinbib,showpacs,superscriptaddress]{revtex4-1}

\usepackage[dvipsnames]{xcolor}
\usepackage{amsmath}
\usepackage{amssymb}
\usepackage{amsfonts}
\usepackage{float}
\usepackage{epsfig}
\usepackage{tikz}
\usepackage[europeanresistors]{circuitikz}
\usepackage{dsfont}
\usepackage{url}
\usepackage{graphicx}
\usepackage[colorlinks]{hyperref}
\hypersetup{%
	plainpages=true,
	breaklinks=true,
	hypertexnames=false,
	pageanchor=true,
	colorlinks=true,
	linkcolor={blue},
	citecolor={red},
	urlcolor={blue},
	anchorcolor={black}
}
\usepackage{bbm}
\usepackage[draft,inline,nomargin,index,marginclue]{fixme}
\fxsetup{status=draft}
\graphicspath{{.}{./figs/}}
\newcommand{\ket}[1]{{\vert #1 \rangle}}

\newcommand{\braket}[2]{\langle{#1}|{#2}\rangle}

\newcommand{\commentD}[1]{{\color{Purple} {\bf DZ:} #1}}
\newcommand{\commentS}[1]{{\color{red} {\bf ST:} #1}}

\begin{document}
\title{Waveguide QED in the Dipole Gauge}
\author{Sergi Terradas-Brians\'o}
\affiliation{Instituto de Nanociencia y Materiales de Arag\'on (INMA), CSIC-Universidad de Zaragoza, Zaragoza 50009, Spain}
\author{Luis Mart\'in-Moreno}
\affiliation{Instituto de Nanociencia y Materiales de Arag\'on (INMA), CSIC-Universidad de Zaragoza, Zaragoza 50009, Spain}
\author{David Zueco}
\affiliation{Instituto de Nanociencia y Materiales de Arag\'on (INMA), CSIC-Universidad de Zaragoza, Zaragoza 50009, Spain}
\date{\today}

\begin{abstract}
In recent studies on ultrastrong coupling between matter and light in cavities, the significance of gauge choice when employing the widely-used two-level approximation has been highlighted.  Expanding upon these investigations, we extend the analysis  to waveguide QED, where we demonstrate that truncations performed in the dipole gauge also yield accurate results. To illustrate this point, we consider the case of a dipole coupled to a cavity array.  Various numerical and analytical techniques have been employed to investigate the low-energy dynamics of the system. Leveraging these theoretical tools, we argue that single photon scattering is an ideal method for investigating gauge-related issues. Our findings reveal two novel effects in the scattering spectra, which cannot be reproduced in a truncated model using the Coulomb gauge. Firstly, the primary resonance is modified due to a Lamb shift contribution. Secondly, we observe asymmetric transmission amplitudes surrounding this resonance, reflecting the asymmetry of the spectral density in this model.  Additionally, we explore other features in the scattering spectra resulting from ultrastrong couplings, such as the emergence of Fano resonances and inelastic channels.  Finally, we propose an experimental test of our ideas in the context of circuit QED.

\end{abstract}
\maketitle

\section{Introduction}%
\label{sec:Intro}

Photons usually interact weakly with matter, which has lead to the adoption of several approximations in quantum optics.  The most common ones include the rotating-wave approximation, truncating the matter description to its lowest energy levels, neglecting the $A^2$ term, and the Markovian approximation for computing spontaneous emission, among others.
However, numerous experiments have demonstrated that discrete quantum emitters can couple to light beyond the limitations of perturbative coupling. This breakthrough has been achieved by coupling these emitters to cavities  \cite{Niemczyk2010,Forn-Diaz2010,Yoshihara2017} and waveguides \cite{Forn-Diaz2017,PuertasMartinez2019,Leger2019, Vrajitoarea2022}. 

It has been demonstrated that, in certain cases, the interaction energy can be comparable to the energies of light and matter.
As a consequence, many approximations break down, indicating that the coupling strength has entered the ultrastrong regime (USC).
One of the primary noticeable effects of entering the USC is the significant emergence of processes that go beyond the interchange of a single photon and matter excitation. %

As a result, the widely used rotating-wave approximation (RWA) for the interaction loses its validity, leading to the renormalization of {the bare emitter} parameters and the emergence of a nontrivial ground state.
Several interesting phenomena have been discussed in relation to the latter, including the possibility of converting virtual photons into real ones through ground state perturbations \cite{Ciuti2005, Stassi2013, QiKai2018, Gheeraert2018, Sanchez-Burillo2019, Cirio2017}, the localization-delocalization transition \cite{Peropadre2013,Shi2018}, and the potential for performing nonlinear optics at the limits of single and zero photons \cite{Sanchez-Burillo2014,Sanchez-Burillo2015,Stassi2017,Kockum2017, Chen2017, Kockum2017a}.
To gain comprehensive insights into light-matter interactions in the USC regime, see \cite{FriskKockum2019,Forn-Diaz2019}.

More recently, it has been discovered that the USC regime introduces additional complications for conventional approaches used to describe light-matter systems.
The commonly employed two-level approximation (TLA) for the matter subsystem has been found to lack gauge invariance in certain descriptions. 
In fact, truncating a momentum-like coupling operator has been shown to cause significant inconsistencies between complete and truncated models \cite{DeBernardis2018a, DiStefano2019, Ashida2022}. 
Instead, employing position-based interactions yields more reliable results when combined with the TLA.
This research area has sparked numerous studies focusing on different approaches to truncating the matter level subsystems \cite{Savasta2021}, and also on the truncation of photonic levels or the number of modes \cite{Taylor2022, SanchezMunoz2018}.
Correctly applying the two-level approximation has significant implications for a wide range of system properties. 
It affects not only the energy levels \cite{DeBernardis2018a}, but also the ground state \cite{Settineri2021}, emission spectra \cite{Mercurio2022, Salmon2022}, spectral density \cite{Salmon2022}, and various other observables \cite{Settineri2021, Salmon2022}.

In this work, we build upon previous approaches to ensure gauge invariance in cavity quantum electrodynamics and apply them to the domain of waveguide QED.
We propose that, for a dipole coupled to a waveguide, employing the dipole gauge is more suitable for matter truncation. 
To illustrate this, we focus on the case of a single {dipole} coupled to a cavity array and conduct both numerical and analytical calculations.
Notably, we demonstrate that scattering experiments are ideal for testing gauge-related issues. 
Specifically, we reveal that the {transmittance minima} exhibits a red-shift as the coupling strength increases, even within the lower range of the ultrastrong coupling regime. 
This red-shift effect arises from the contribution of Lamb shifts, which contrasts with the constant resonance observed when truncation is performed in the Coulomb gauge.
This key result highlights a qualitative distinction between truncation approaches carried out in two different gauges.
Furthermore, our analysis reveals that including counterrotating terms enables the occurrence of Fano resonances and inelastic scattering processes.

The rest of the manuscript is organized as follows. 
In Sec. \ref{sec:GI_wQED}, we present a general gauge-invariant description of a single particle interacting with the electromagnetic field. 
We apply this description to the case of waveguide QED and particularize it to a cavity array waveguide.
Section \ref{sec:scattering} introduces the framework used to address the scattering processes in these systems and discusses how gauge invariance affects their description.
The different theoretical methods employed in the scattering computations are described in Sec. \ref{sec:methods}. 
In section \ref{sec:results}, we present and analyse the numerical results for the scattering spectra both within and beyond the rotating-wave approximation.
Sec. \ref{sec:implementation} proposes an implementation of this setup in a circuit QED platform. 
Finally, we conclude in section \ref{sec:Conclusions}.
The manuscript also includes four appendices:
Appendix \ref{app:weak} analyses gauge invariance in the weak coupling limit.
Appendix \ref{sec:app_SelfEn} provides the analytical computation of the self-energy in the RWA.
Appendix \ref{sec:app_CoulombT} discusses the convergence of Coulomb gauge and dipole gauge simulations.
Appendix \ref{sec:app_ContLim} deals with the continuous limit of the dipole gauge model.

\section{Gauge invariant formulation of waveguide QED}
%
%
%
%

\label{sec:GI_wQED}

\subsection{Preliminaries}
%
%
%
%
\label{sec:wQED_Pre}

In this section, we present a formalism that describes the interaction between a single emitter and the electromagnetic field. A convenient approach to introducing gauge-invariant light-matter systems can be found in Refs. \cite{DiStefano2019, Savasta2021, Dmytruk2021}.  
In the Coulomb gauge, the Hamiltonian is expressed as:
\begin{equation}
\label{Hc}
    H_C = H_{\rm ph} + U^\dagger H_{\rm m} U \; .
\end{equation}
Here, $H_{\rm ph}$ represents the quantized Hamiltonian for the electromagnetic field, which will be specified below, and $H_{\rm m}$ refers to the matter Hamiltonian.
The unitary operator $U$, which represents a gauge transformation itself, is given by
\begin{equation}
\label{U}
    U = e^{i q {\bf A(x)} \cdot {\bf x}/\hbar}
    \; ,
\end{equation}
where ${\bf A(x)}$ is the vector potential and $q$ denotes the {emitter} charge.
The variable ${\bf x}$ corresponds to the position operator of the {emitter}. It is worth noting that ${\bf A (x)}$ can explicitly depend on this position.

Given $H_{\rm m} = {\bf p}^2/2m + V ({\bf x})$,  Eqs. \eqref{Hc} and \eqref{U} result in the minimal coupling Hamiltonian
\begin{equation}
\label{minimal}
H_C = \frac{1}{2m}  \Big ( {\bf p}- q {\bf A} \Big )^2+ V ({\bf x}) \; .    
\end{equation}

This way of expressing $H_C$ also offers a straightforward understanding of the dipole gauge. 
In fact, the Hamiltonian in the dipole gauge can be represented by an gauge transformation, equivalent to the Power-Zienau-Woolley (PZW) one, as [Cf. Eq. \eqref{Hc}],
\begin{equation}
\label{HD}
    H_D = H_{\rm m} + U H_{\rm ph} U^\dagger
    \; .
\end{equation}
Within this formulation, it is evident that $H_D = U H_C U^\dagger$, ensuring gauge invariance.
Furthermore, the light-matter coupling transforms the \emph{bare} matter Hamiltonian in the Coulomb gauge, while in the dipole gauge, it transforms the \emph{free} electromagnetic field Hamiltonian.
Up to this point, there should be no ambiguity about working in one gauge or the other.
However, in practical calculations, we often need to make approximations in the Hamiltonians, especially when focusing on low-energy dynamics.
In such cases, we typically truncate the matter Hamiltonian to some minimum energy states, using methods like the two-level approximation or the single band limit, among others.
The problem of gauge ambiguities arises when making these truncations, since {applying approximations on the matter subsystem}
in one gauge may lead to different results than in another. 
This discrepancy can be understood by examining Eqs. \eqref{Hc} and \eqref{HD}: matter and photonic operators are not the same in $H_C$ as in $H_D$ [cf. Eq. \eqref{U}]. 
For instance, in single-mode cavity QED, where ${\bf A (x)} = {\bf A_0 (x_0)} (a + a^\dagger)$, it has been argued that the correct starting point for performing the truncation is $H_D$, while truncation in $H_C$ yields incorrect results in the USC regime. 
This is because in the dipole gauge, the coupling is ${\bf E (x) \cdot x}$ (${E} ({\bf x})$ is the electric field), while in the Coulomb gauge, it goes as ${\bf A (x) \cdot p}$. 
As {the emitter} wave functions are typically localized in space, the momentum-${\bf p}$ matrix elements can not be neglected even between well-separated (in energy) eigenstates \cite{DeBernardis2018a}.
The same is expected to occur in waveguide QED, as discussed below.

\subsection{Waveguide QED Hamiltonian}
%
%
%
%
\label{sec:wQED_H}

EM quantization is convenient in the Coulomb gauge. In what follows, we will assume that the \emph{position} of the emitters is fixed. 
This is the case in the majority of situations and facilitates the discussion. 
To be more precise, we will assume that the emitter position can be written as ${\bf x} = {\bf x_0} + \delta {\bf x}$ (${\bf x_0}$ is not longer an operator but a vector position). 
For all the relevant energy scales ${\bf A (x)} \cong {\bf A (x_0)}$, which is the long-wavelength approximation.
In this scenario, ${\bf A}$ acts only on the Hilbert space of the photons \cite[Sect. 3]{RomanRoche2022}.
\begin{equation}
\label{ACQ}
    {\bf A}_\perp ({\bf x_0}) =
    \frac{1}{\sqrt{L}} \sum_k   \Big (  { \boldsymbol \lambda}_k (x_0, y_0) 
    \; a_k e^{i k z_0} + {\rm h.c.} \Big ) \; .
\end{equation}
We have added the suffix $\perp$ to emphasize that the potential vector has only transverse components in the Coulomb gauge.
In this work, we focus on waveguide QED, which fixes a propagation direction, denoted as $z$ (cf. the exponentials and the scalar character for the wavevector $k$). 
Additionally, this  constrains $ { \boldsymbol \lambda}_k$ to the $xy$-plane, which can be expressed as:
\begin{equation}
   { \boldsymbol \lambda}_k \equiv \sqrt{\frac{\hbar}{2 \omega_{\bf k} \epsilon_0}}
    {\bf u}_{\bf k} \; ,
\end{equation}
with ${\bf u}_k$ normalized functions, satisfying $\int_{\mathbb R_2}  dx dy \, |{\bf u}_k |^2 = 1$,  and providing the space dependence of ${\bf A}$ around the waveguide.
%
The waveguide is assumed to be surrounded by vacuum, hence the appearance of $\epsilon_0$ above. The frequencies of the different waveguide modes are denoted as $\omega_k$:
\begin{equation}
    H_{\rm ph} =  \sum_k \hbar\omega_k a_k^\dagger a_k \; . 
\end{equation}
Up to this point, we have assumed a waveguide of length $L$, using a discrete number of modes $k = m \times \pi / L $, where $m=0,\pm 1, ...$. However,  the continuum limit can be obtained by taking $L \to \infty$.

At this point, it becomes evident that continuing to use the Coulomb gauge comes with certain obstacles.
The most evident is the fact that the minimal coupling, introduces the ${\bf A}^2$-term  which, after Eq. \eqref{ACQ}, couples all the waveguide modes, resulting in terms like $a_k a_{k^\prime} + a_k^\dagger a_{k^\prime} + {\rm h.c.}$.
As a consequence, the photonic part should be diagonalized using a Bogolioubov transformation.

Alternatively, we can switch to the dipole gauge \eqref{HD}. 
To achieve this, we just need to know how $a_k$ transforms under $U$ in \eqref{U},
\begin{equation}
\label{atoD}
    U a_k U^\dagger = a_k - i  q   \frac{1}{\sqrt{L}} { \boldsymbol \lambda}_k e^{-i k z_0} \cdot {\bf x}
\end{equation}
Therefore, the waveguide QED Hamiltonian in the dipole gauge can be written as:
\begin{widetext}
\begin{equation}
\label{HDsb}
    H_D = H_{\rm m} + \sum_k\hbar \omega_k a_k^\dagger a_k 
    - i q \frac{1}{\sqrt{L}} {\bf x} \sum_k  \Big ( {\boldsymbol \lambda}_k \hbar\omega_k e^{i k z_0} a_k - {\rm h.c.} \Big ) +
     \frac{q^2}{L }  \sum_k  \hbar\omega_k  \Big | 
    {\boldsymbol \lambda}_k  e^{i k z_0} \cdot {\bf x} \Big |^2.
\end{equation}
\end{widetext}

The last term in Eq. \eqref{HDsb}  can be absorbed into $H_{\rm m}$, resulting in a modified matter Hamiltonian ${H}^\prime_{\rm m}= H_{\rm m} +\frac{q^2}{L }  \sum_k \hbar \omega_k  \Big | {\boldsymbol \lambda}_k  e^{i k z_0} \cdot {\bf x} \Big |^2 $.
If ${H}^\prime_{\rm m}$ can be described using its two lowest states, say $\{ |{0}^\prime \rangle, |{1}^\prime \rangle \}$, $H_D$ can be expressed in the form of a spin-boson model:
\begin{equation}
\label{HDtls}
    {\mathcal H}_D = \frac{\hbar{\Delta}^\prime}{2} \sigma_z +  \sum_k \hbar\omega_k a_k^\dagger a_k + \sigma_x \sum_k \Big (\hbar g_k a_k + {\rm h.c.} \Big )
\end{equation}
where
\begin{equation}
    \label{eq:gk_gen}
    g_k = \frac{\omega_k}{\sqrt{L}} \langle {0}^\prime | {\bf d} | {1}^\prime \rangle\cdot {\boldsymbol \lambda}_k,
\end{equation}
with $\hbar {\Delta}^\prime$ being the transition energy between the two states and ${\bf d} = q {\bf x}$.
Here and throughout the text, we use the notation $\mathcal{H}_D$ to distinguish the truncated Hamiltonian in Eq. \eqref{HDtls} from the full model $H_D$ in \eqref{HDsb}.
%
It is customary to define the spectral density for spin-boson models as:
\begin{equation}
    \label{eq:SpectrDensDef}
    J_D (\omega) = 2 \pi \sum_k |g_k|^2 \delta (\omega - \omega_k) \; .
\end{equation}
Here, we have used the suffix $D$ to emphasize that the spectral density is gauge-dependent since, it depends on the 
light-matter coupling 
\eqref{eq:gk_gen}.

As mentioned earlier, the ${\bf A}^2$-term in the Coulomb gauge prevents us from explicitly expressing the Hamiltonian in the form of \eqref{HDsb} and/or \eqref{HDtls}.  
However, at low and intermediate couplings, the ${\bf A}^2$ is usually neglected. By doing so, one can obtain a spin-boson model, but with a different transition frequency $\Delta$ (the last term of \eqref{HDsb} has been absorbed into $H_{\rm m}$) and different a spectral density. 
Since the spin-boson model is determined by its the spectral density, the Coulomb and dipole gauges are not equivalent after truncation, similar to the case in cavity QED.
Despite this, it is possible to show that,
\begin{equation}
    \label{eq:LimitSpectrDens}
    \lim_{|{\boldsymbol \lambda}_k | \to 0} J_C (\Delta) \to J_D (\Delta) 
\end{equation}
The proof is provided in Appendix \ref{app:weak}.

Thus, truncation can be safely done in both gauges in the usual scenario of waveguide QED, in which the light-matter coupling is \emph{small}. 
However, entering the USC regime requires more caution, and as we will discuss in detail later, the dipole gauge proves to be quite convenient. In addition, we will explore the physical consequences of working in the USC regime and how they manifest in standard experiments, such as single photon scattering.

\subsection{The Cavity array case}
%
%
%
%
\label{sec:wQED_CA}

\begin{figure}[t]
    \centering
    \includegraphics[width = \columnwidth]{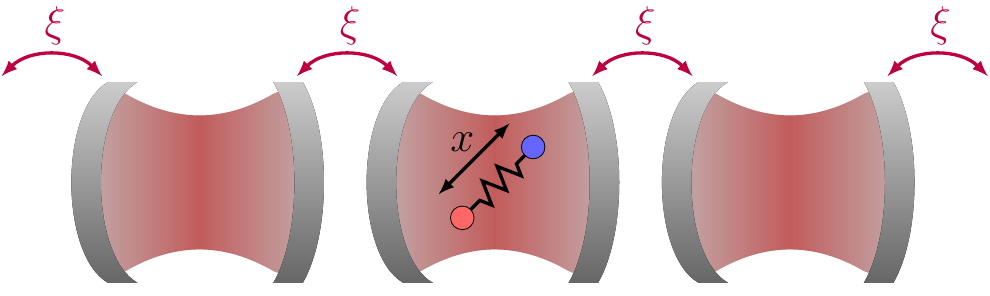}
    \caption{Schematic representation of a one-dimensional array of coupled cavities via the hopping parameter $\xi$. The dipole lies within the centre cavity with a relative distance between its charges $x$. }
    \label{fig:SchemeCavArray}
\end{figure}

Up to this point, we have not specified the particular waveguide or emitter we are considering. Moving forward, we will analyze a model that allows for both exact numerical treatments and analytical estimates.
To describe the emitter, we will assume it to be spherically symmetric {dipole}, which enables us to use a one-dimensional model with position $x$ and momentum $p$ operators to represent its Hamiltonian.
On the other hand, we will focus on a cavity array for the waveguide, where the emitter is coupled to a single cavity (specifically, the $n=0$ cavity). 
We will also assume that the cavities in the array are single-mode, with the vector potential aligned with the dipole in the Coulomb gauge. This setup ensures that the model remains fully one-dimensional. 
A schematic representation of this system is shown in Fig. \ref{fig:SchemeCavArray}.
As we will discuss, these simplifications can be justified within the circuit QED architecture, where the USC regime has been reached, and our proposed ideas may be implemented.
Then, the resulting \emph{full} light-matter Hamiltonian in the Coulomb gauge can be expressed as:

\begin{align}
    \label{eq:HCoulomb}
     H_{C} &= \frac{\left[p-q {A}_0(a_0 +a^\dagger_0)\right]^2}{2m} + V(x) \nonumber\\
     &+\hbar\omega_c \sum_n^N a^\dagger_n a_n+ \hbar\xi\sum_n^N (a_n a^\dagger_{n+1}+ a^\dagger_n a_{n+1}) 
\end{align}
In this model, we consider identical $N$ cavities, each with a resonance frequency $\omega_c$. 
The cavities are coupled to their left and right neighbors with a strength $\xi$ in a tight-binding fashion (last term of equation \eqref{eq:HCoulomb}).
Regarding the dipole, we describe it using a double-well potential:
\begin{equation}
    V(x) = -\mu \frac{x^2}{2}  + \lambda \frac{x^4}{4} .
\end{equation}

Moving to the momenta space  $a_k = 1/\sqrt{N}\sum_n a_n e^{ikn}$ and transforming to the dipole
gauge following Eqs. \eqref{HD}, \eqref{atoD}, we obtain the waveguide Hamiltonian with $H_{\rm m }^\prime = p^2/2m + V(x) +\hbar\omega_c q^2 {A}^2_0 x^2$ [Cf. Eqs. \eqref{HDsb} and \eqref{eq:HCoulomb}] and,
\begin{equation}
    \lambda_k =  A_0 
\end{equation}
Our simulations consider both the non-truncated (which we call the full model) and the truncated version where only the two lowest states, $\{ |0^\prime \rangle, |1^\prime \rangle\} $ of $H_{\rm m}^\prime$ are retained.
In the latter case, the model to consider is the spin-boson one \eqref{HDtls} with,
\begin{align}
    \label{eq:gk}
    g_k = \frac{\omega_c q A_0 | \langle 1^\prime | {x} | 0^\prime \rangle |}{\sqrt{N}}
    \frac{\omega_k}{\omega_c},
\end{align}
then from Eq. \eqref{eq:SpectrDensDef}
\begin{align}
    \label{eq:JD}
    J_D(\omega) = 
    \frac{2 g^2 }{\sqrt{4\xi^2 - (\omega - \omega_c)^2}} \frac{\omega^2}{\omega^2_c},
\end{align}
to quantify the coupling regime of the system in the truncated dipole gauge, we define the coupling strength $g = q{A}_0 \omega_c | \langle 1^\prime | x | 0^\prime \rangle|$.

Moreover, we find it useful to express the dipole gauge Hamiltonian in position space, as we will employ this representation in our scattering calculations. 
Using $a_n = 1/\sqrt{N}  \sum_k a_k e^{-ikn}$, it yields,
\begin{widetext}
\begin{align}
    \label{eq:HDipole-position}
    H_D = &  H_m^\prime +\hbar\omega_c \sum_n a^\dagger_n a_n +
     \hbar\xi\sum_n (a^\dagger_n a_{n+1}+ {\rm h.c.})
     -i \hbar\omega_c q {A}_0(a^\dagger_0-a_0)x
    -i \hbar\xi q{A}_0\left[(a^\dagger_1-a_1)+(a^\dagger_{-1}-a_{-1})\right]x.
\end{align}
\end{widetext}
%
%
The gauge transformation intertwines light and matter; hence, in this new gauge, the dipole also couples to the adjacent cavities $n=-1,1$, which, in the Coulomb gauge, are the ones coupled to cavity $n=0$ via the hopping term.

\subsection{Truncation of matter levels in both gauges}

%
%
%
%
\label{sec:wQED_Trunc}

\begin{figure}
    \centering
    \includegraphics[width = \columnwidth]{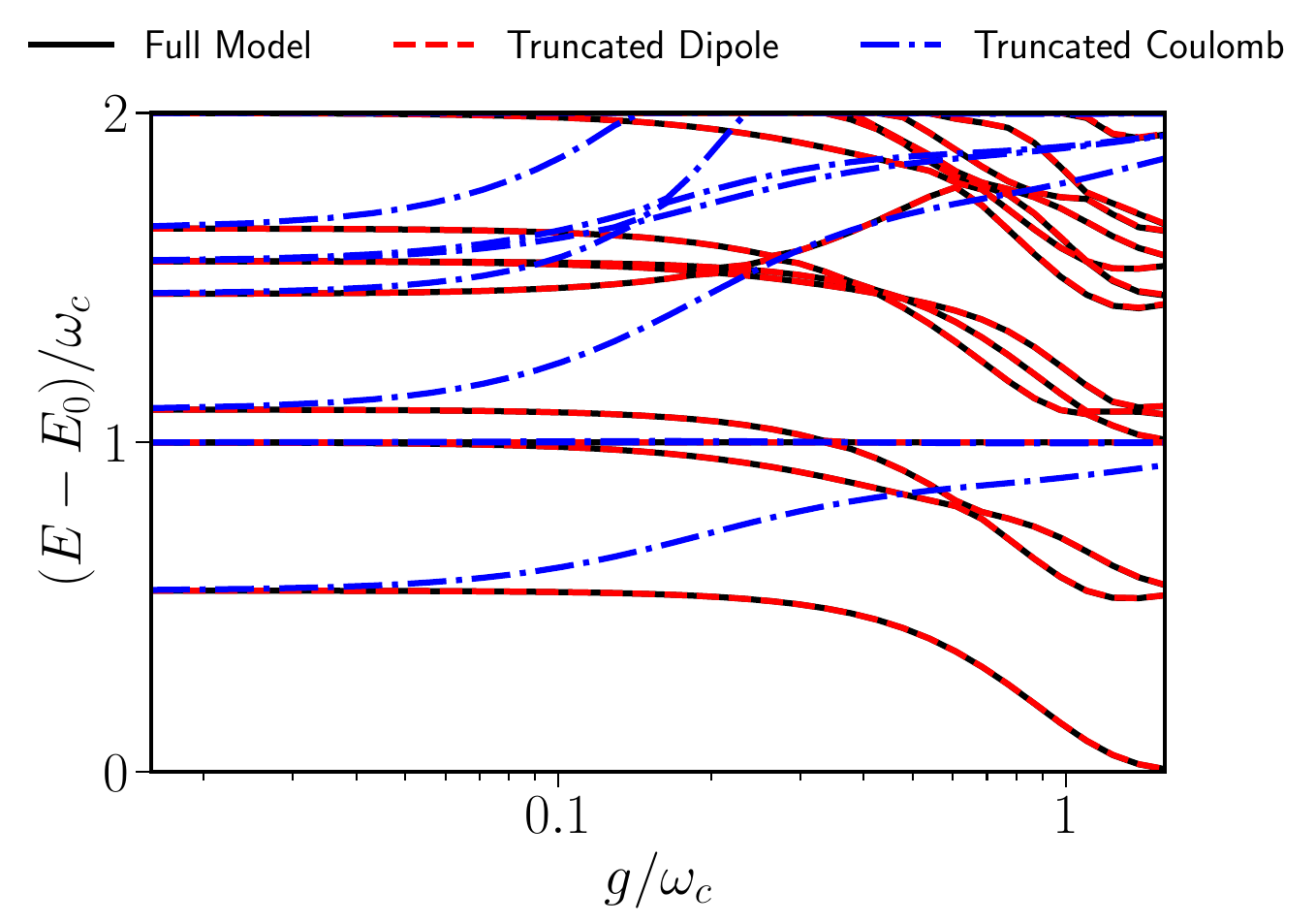}
    \caption{
    The energy spectra of an array of three cavities with the central one coupled to the dipole, are shown for the full model (solid black line), the truncated dipole gauge (dashed red line) and the truncated Coulomb gauge (dashed-dotted blue line).
    In this plot,  we choose $\beta = 3.8$ and $E_{dip}/(\hbar\omega_c) = 63.812$ such that at zero coupling, there is a resonance between the cavity energy $\omega_c$ and the first bare dipole transition $\Delta$.
    The hopping parameter is given by $\xi/\omega_c = -1/\pi$.
    For the full model diagonalization, we consider 18 dipole levels and 18 photonic levels in each of the three cavities, while the truncated cases use the same number of photonic excitations and 2 dipole levels. }
    \label{fig:Spectra}
\end{figure}

In general, Hamiltonians \eqref{HDsb}, \eqref{eq:HCoulomb} or even the truncated version \eqref{HDtls} are non-integrable. 
Besides, the numerical calculation of the spectrum is rather challenging.
Hence, performing a benchmark of the truncation in different gauges with the full model is not viable.
%
{To illustrate how the truncation affects gauge invariance, we present the eigenenergies of a minimial system comprising a dipole coupled to three cavities. 
These are obtained in a gauge invariant non-truncated representation and within the two-level approximation in both the Coulomb and dipole gauges in Fig. \ref{fig:Spectra}.}

Refs. \cite{DeBernardis2018a, DiStefano2019} introduce a convenient dimensionless notation for the matter Hamiltonian, which we also follow in the diagonalization procedure to obtain the spectra of Fig. \ref{fig:Spectra}.
%
By defining a length scale $x_0 = (\hbar^2/(\lambda m))^{1/6}$, we work with the dimensionless variable $z = x/x_0$, allowing us to rewrite the bare matter Hamiltonian $H_m$ as:
\begin{align}
    \label{eq:H_dip}
    H_{m} =E_{d}\left[ \frac{p^2_z}{2} - \frac{\beta z^2}{2} + \frac{z^4}{4}\right],
\end{align}
where $\beta = m\mu x^4_0/\hbar^2$, $p_z = - i\hbar\partial/\partial z$.

Fig. \ref{fig:Spectra} demonstrates that, similar to the single cavity QED case \cite{DeBernardis2018a, DiStefano2019}, the truncation fails in the Coulomb gauge, while it accurately matches the full model in the dipole gauge.
This observation highlights that the truncation in the dipole gauge, with an interaction described by the position operator, preserves the energy levels of the Hamiltonian.
In contrast, calculations within the Coulomb gauge are profoundly affected by the truncation, which is consistent with the results obtained the single cavity scenario \cite{DeBernardis2018a, DiStefano2019}.%

\section{Scattering: setting the problem}
\label{sec:scattering}
%
%
%
%
A caveat in USC is that the distinction between ``light" and ``matter" subsystems becomes less clear.
This is not only due to the creation of strongly correlated light-matter states but also because the definition of these subsystems is gauge-dependent.
While physical observables remain gauge-invariant, the choice of matter or light observables can be ambiguous.
%
%
To overcome these subtleties and provide \emph {clear and measurable signatures of truncation issues in different gauges}, this paper focuses on studying scattering phenomena.
Scattering serves as an ideal testbed for highlighting gauge-related problems.
The scattering problem can be simplified as follows. 
The input state, representing our initial condition, is chosen to be the non-normalized quantum state,
\begin{equation}
\label{eq:psi_in}|\Psi_{in}\rangle = 
(a_\phi^\dagger)^N|\mathrm{GS}\rangle,\qquad a_\phi^\dagger = \sum_x \phi^{\rm in}_x a_x^\dagger\, ,
\end{equation}
where $|\mathrm{GS}\rangle$ is the ground state of the system 
 and $\phi_x^{\rm in}$ is a Gaussian wavepacket centred in $x_{\rm in}$
 with spatial width $\theta$,
\begin{equation}
\label{eq:wp}
\phi_x^{\rm in}=\exp\left(-\frac{(x-x_{\rm
      in})^2}{2 \theta ^2}+ik_{\rm in}x\right)
\; .
\end{equation}
Typically, we consider $x_{\rm in}$ located on the left-hand side of the scatterer, with the wavepacket moves to the right towards it.
The wavepacket (\ref{eq:wp}) is exponentially localized around $k_{\rm in}$, with a width of approximately $\theta^{-1}$.

The wave packet evolves in time as
\begin{equation}
| \Psi (t) \rangle = U(t, 0) |
\Psi_{\rm in} \rangle = {\rm e}^{-i H_{\rm tot} t} | \Psi_{\rm in} \rangle 
\; .
\end{equation}
A final time $t_\mathrm{out}$ is chosen, which must be sufficiently
large to allow the photons to move freely along the waveguide after interacting with the scatterers.
The evolution is then described by the $S$-matrix, defined as:
\begin{equation}
\label{eq:psi_out}|\Psi_{\rm out}\rangle = S|\Psi_{\rm in}\rangle\, .
\end{equation}
The scattering matrix is characterized by its momentum components:
\begin{equation}
\label{eq:Spk}
S_{p_1  ...  p_{N^{\prime}} , \, k_1 ...  k_N}
=
\langle \mathrm{GS}  | a_{p_1} ... a_{p_{N^\prime}}  \, S  \, a_{k_1}^\dagger
... a_{k_{N}}^\dagger  | \mathrm{GS} \rangle
\; .
\end{equation}
Some comments are pertinent here.  
The ground state wavefunction $|\mathrm{GS}\rangle$ appears in the definition of $S$.
In the ultrastrong coupling regime,  as discussed before, the ground state differs from the vacuum state and contains a non-zero number of excitations \cite{Sanchez-Burillo2019}.
In this regime, the number of excitations is not conserved, so $N^\prime \neq N$ in general in the Eq. \eqref{eq:Spk}.
However, we can expect some simplifications to occur.
We specialized our discussion to single photon wavepackets, searching for computational simplicity, and also because in the experiments, when using low-power coherent classical input/output field states, the transmittance amplitudes coincide with the single photon scattering amplitudes.
Moreover, it has been shown \cite{SanchezBurillo2018} that for wavepackets far away from the scatterer, even in the ultrastrong coupling regime, we can approximate $a_\phi^N |\mathrm{GS} \rangle \cong a_\phi^N |{\rm vac}\rangle$, where $|{\rm vac} \rangle$ represents the trivial vacuum of the waveguide with $a_n | {\rm vac} \rangle = 0$ for all $n$.
Lastly, it has been numerically tested that the probability of having more of one photon in the output field is negligible \cite{Sanchez-Burillo2014, Sanchez-Burillo2015}. 
As a consequence, the single photon amplitudes can be related to the number of photons as follows:
\begin{equation}
    \label{eq:transmitt}
    {\frac{\langle {\Psi_{\rm out}} |
    a_k^\dagger a_k | \Psi_{\rm out} \rangle}
    {\langle {\Psi_{\rm in}} |
    a_k^\dagger a_k | \Psi_{\rm in} \rangle}}
    = |S_{kk}|^2 \equiv t_k.
\end{equation}
This equation defines the {\emph{transmission amplitude}} $t_k$, which is a key quantity in this work.
 
Within this framework, it is easy to understand how scattering is free from ambiguities in the abovementioned sense. 
Of course, being an observable, the transmission amplitude is gauge invariant. 
What makes scattering ``special" is that both the input and output fields have support only on regions well separated from the scatterer. 
Consequently, they can be considered as free photon wavepackets created over the QED vacuum of the waveguide, which remains the same in all the gauges: $U | \Psi_{\rm in} \rangle = | \Psi_{\rm in} \rangle$ (the same with $|\Psi_{\rm out} \rangle$).
Therefore, one can start with the same initial conditions in both gauges, allow the system to evolve, compute the amplitude and compare the results after performing truncations in both the dipole and Coulomb gauges.

\section{Theoretical methods.}
\label{sec:methods}

%
%
%
%

Let us outline the theoretical methods we employed to compute the scattering spectra in the USC regime. 
Our approaches involve a Polaron-like transformation, numerical simulations based on Matrix Product States, and matching techniques. 
The readers already familiar with these methods or not interested in technical details, should better proceed to the next section.
\subsection{Polaron picture: effective single excitation dynamics}
\label{sec:M_polaron}

%
%
%
%

The polaron formalism provides an effective description of the low-energy sector of spin-boson models such as \eqref{HDtls}.
This approach is based on applying a unitary transform that disentangles the light and matter subsystems.
\begin{equation}
    \label{eq:Upol}
    U_\mathrm{P} = \exp\left(-\sigma_x \sum_k f^{*}_k a_k - f_k a^\dagger_k \right),
\end{equation}
where $f_k$ represents a set of variational parameters.
These parameters are found by minimising the ground state energy, which is given by the state with zero excitations $\ket{\Psi^\mathrm{GS}_\mathrm{P}} = \ket{0}\otimes\ket{0_k}$, see \emph{e.g.} \cite{Silbey1983, Bera2014, Diaz-Camacho2016, Zueco2018}.

After applying \eqref{eq:Upol}, the effective Hamiltonian in the polaron picture is given by
\begin{align}
    \label{eq:HPol}
    \mathcal{H}_P &= \sum_k \omega_k a^\dagger_k a_k + \frac{\Delta_r}{2}\sigma_z  + 2\Delta_r\sum_k (f_k a_k \sigma^{+} + \mathrm{h.c.}) \nonumber
    \\
    &-2\Delta_r \sigma_z \sum_{k,p} f_k f^{*}_p a^\dagger_k a_p,
\end{align}
where 
\begin{equation}
\label{Deltar}
 \Delta_r = \exp\left(-2\sum_k \lvert f_k \rvert^2\right),
\end{equation}
and 
\begin{equation}
\label{fk}
   f_k =\frac{ g_k}{(\Delta_r + \omega_k) }.
\end{equation}
Here and throughout the manuscript, we use the convention $\hbar = 1$ for simplicity.
The variational parameters $f_k$ and $\Delta_r$ are obtained self-consistently from Eqs. \eqref{Deltar} and \eqref{fk}.
The renormalized frequency $\Delta_r$ is a well-known result from the spin-boson model.
It tends to vanish as the coupling increases, eventually leading to the localization-delocalization quantum phase transition depending on $J(\omega)$ \cite{Leggett1987}.
The main advantage of the effective Hamiltonian \eqref{eq:HPol} is that it conserves the number of excitations, making the single-particle dynamics relatively straightforward.

\subsection{Matrix product states}

%
%
%
%

\label{sec:M_MPS}

Tensor networks have proven to be effective tools for numerically simulating light-matter systems in the ultrastrong coupling regime \cite{Peropadre2013, Sanchez-Burillo2015, Shi2018}.
Specifically, for our one-dimensional chain with nearest-neighbor interactions described by \eqref{eq:HDipole-position}, we utilize matrix product states in conjunction with a time evolution block decimation (TEBD) algorithm \cite{Garcia-Ripoll2006} to simulate wavepacket scattering, as discussed in Section \ref{sec:scattering}.
By simulating the scattering process for a sufficiently long duration $t_{out}$, we can obtain the outgoing eigenstate \eqref{eq:psi_out}, as explained in the previous section.
Furthermore, with the matrix product state representation of $\ket{\Psi_{\rm in}}$ and $\ket{\Psi_{\rm out}}$, the computation of the {transmission amplitude} transmittance described by Eq. \eqref{eq:transmitt} can be efficiently implemented.


\subsection{Matching techniques}
\label{sec:M_Matching}

%
%
%
%

\begin{figure}[h]
    \centering
    \includegraphics[width = \columnwidth]{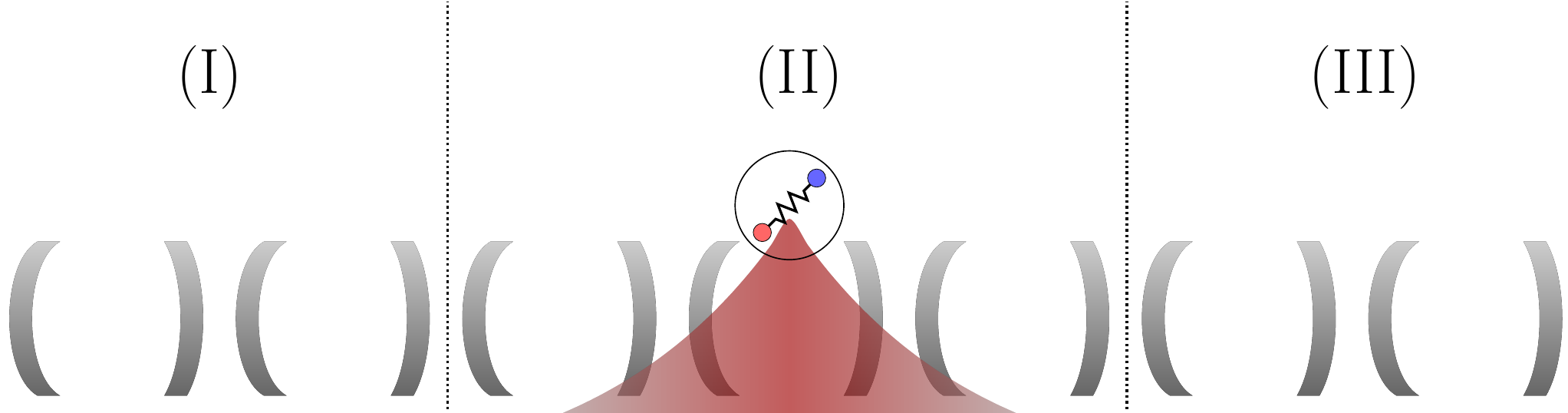}
    \caption{The coupled waveguide is divided into three regions (I), (II), and (III), which are employed in computing the transmission via matching methods. Regions (I) and (III) are chosen to have negligible ground-state photonic population, even in the ultrastrong coupling regime.}
    \label{fig:matching_regions}
\end{figure}

Lastly, we also use a matching technique to obtain the transmittance and reflectance amplitudes.
The fundamental concept in this framework involves dividing the coupled waveguide into three distinct regions, as depicted in Figure \ref{fig:matching_regions}.
The central region (II) encompasses the dipole, the central cavity, and a sufficient number of additional cavities to contain the virtual photonic excitations resulting from the dressing of the dipole in the ultrastrong coupling regime.
Ground state photons exhibit exponential localization around the emitter, with $\langle a_n^\dagger a_n \rangle_{\rm gs} \sim e^{-\kappa_{\rm GS} |n|/2}$, where $\kappa_{\rm GS}$ denotes the localization length (which has been calculated in \cite{Roman-Roche2020} but is irrelevant for our purposes).
Consequently, the cavity regions (I) and (III) can be considered, to a good approximation, to not have photons in the ground state.
In other words, in regions (I) and (III), we assume that $\langle a_n^\dagger a_n \rangle_{\rm GS}=0$.

Bringing everything together, in regions (I) and (II), we describe single-photon transport using the following state
\begin{equation}
    \label{eq:AnsatzMatching}
    \ket{\Psi} = \sum_{n,\alpha} \phi_{n,\alpha} a^\dagger_n \ket{0_{ph},\alpha_{sc}} + \sum_\alpha f_\alpha \ket{0_{ph},\alpha_{sc}} \; .
\end{equation}
Here, $\ket{0_{ph}, \alpha_{sc}}$ represents the state with zero excitations in regions (I) and (III). 
The label $\alpha$ denotes the eigenstates in the region (II), which can be computed numerically. 
For the parameters considered in this work, we have found that up to 5 or 7 cavities in region (II) are sufficient to obtain an accurate description of the transmittance.

The quantity $\phi_{n,\alpha}$ denotes the amplitude of having a photon in the n-th cavity while the scatterer remains in state $\alpha$:
\begin{equation}
\label{eq:AnsatzMatching2}
\phi_{n,\alpha}(k) =
\begin{cases}
 e^{ik n} + r_{k,\alpha}e^{-ik n}, & \mathrm{(I)}\\
 t_{k, \alpha}e^{ik n}, & \mathrm{(III)}
\end{cases}
\; .
\end{equation}

By employing the ansatz \eqref{eq:AnsatzMatching}, the time-independent Schr\"{o}dinger equation can be solved, resulting in the determination of the transmittance and reflectance amplitudes.


\section{Results}

%
%
%
%

\label{sec:results}

The complete transmittance spectrum is quite intricate, as we will soon discover. 
To better understand the gauge issues and the impact of truncation in different gauges, we will initially focus on calculating the scattering coefficients within the rotating-wave approximation.
This approach simplifies single photon scattering and allows for fully analytical solutions.
Subsequently, we will proceed to solve the full model and comprehensively discuss the complete transmittance spectrum.

\subsection{Transmission under the RWA approximation}
%
%
%
%

\begin{figure}[H]
    \centering
    \includegraphics[width = \columnwidth]{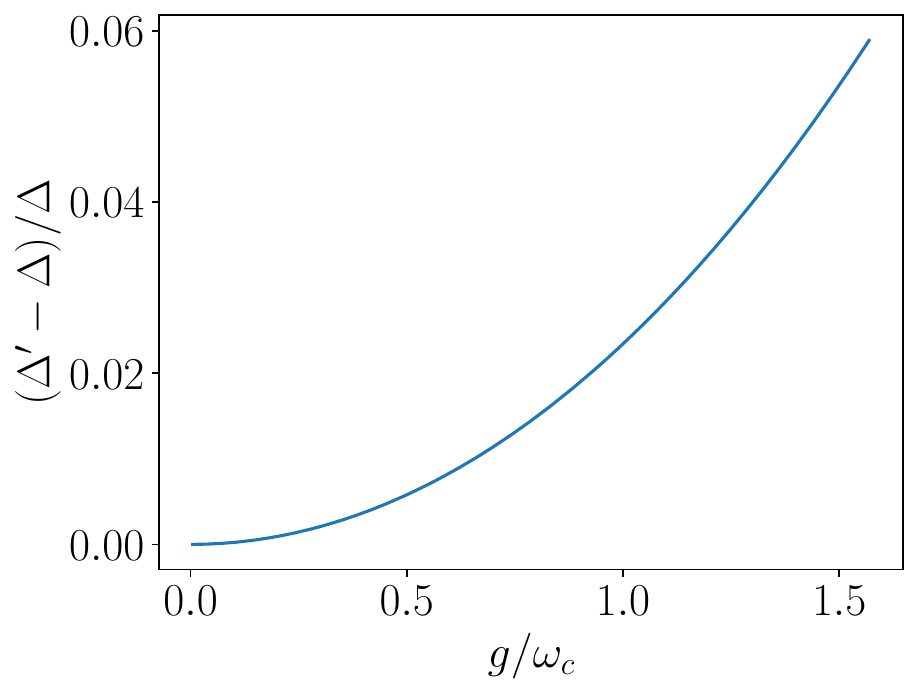}
    \caption{Renormalization of the first dipole transition energy in the dipole gauge $\Delta^\prime$ with increasing coupling, presented in units of the bare energy. The parameters used are the same as in Fig. \ref{fig:Spectra}.}
    \label{fig:Deltacomp}
\end{figure}

In order to isolate the effects of the truncation, we will apply the following approximations.
We  apply the two-level truncation and the RWA such that $\mathcal{H}_D =\mathcal{H}_0 + \mathcal{H}_I$, with
\begin{equation}
\label{eq:H0_RWA}
\mathcal{H}_0 
=
\Delta \sigma^+ \sigma^- 
+
\omega_c \sum_n a_n^\dagger a_n
+
\xi\sum_n (a^\dagger_n a_{n+1}+{\rm h.c.})
\end{equation}
and 
\begin{equation}
\label{eq:H1_RWA}
    \mathcal{H}_I = g (i \sigma^{-}  a_0^\dagger + {\rm h.c.})
    +\frac{\xi g}{\omega_c}\left (  \sigma^- (i a_1^\dagger+ i a_{-1}^\dagger ) + {\rm h.c.} \right ) 
\end{equation}
$\sigma^{\pm}$ are the two level system ladder operators and $g= \omega_c q{A}_0 \lvert \braket{0^\prime|x}{1^\prime} \rvert$. 
We emphasize that the rotating-wave approximation is valid only in the weak coupling regime, but here we extend our computations to larger couplings to compare the RWA results directly with those obtained from the full model, thereby revealing the role of the counter-rotating terms.
\\
%
%
%
%
{Furthermore, we neglect the terms of second order in the couplings, specifically the $x^2$ term in the dipole Hamiltonian.
As mentioned, this term introduces a dependency on the coupling strength $g$ to the TLS transition $\Delta^\prime$ in Eq. \eqref{HDtls}. 
However, as shown in Figure \ref{fig:Deltacomp}, $\Delta^\prime$ increases with $g$ but we note that this change is relatively small compared to the other effects discussed in this section.
Therefore, for the RWA calculations, we will not consider this effect to gain a qualitative understanding.
}

The rotating-wave approximation offers several advantages. First, the ground state becomes trivial, with $|\mathrm{GS} \rangle = |0 \rangle$, implying zero excitations in both the waveguide and the dipole.
Moreover, within the RWA, we can work within the single excitation manifold since the number of excitations is conserved, i.e., $[\mathcal{H}_D, N]= 0$, where $N = \sum_n a^\dagger_n a_n + \sigma^{+}\sigma^{-}$. 
In this case, the matching method introduced in Sec. \ref{sec:M_Matching} becomes exact using the general \emph{ansatz} for a quantum state in the single excitation manifold.
\begin{equation}
\label{eq:SingleExcAntz}
\ket{\psi} = \sum_n \phi_n(k) a^\dagger_n \ket{0} + \phi_q \sigma^{+} \ket{0} \; .
\end{equation}

The eigenvalue problem, $\mathcal{H}_D \ket{\psi} = E\ket{\psi}$, leads to a dipole excited state amplitude given by:
\begin{equation}
    \phi_q = \frac{g}{E-\Delta} \left[\phi_0(k) + \frac{\xi}{\omega_c}(\phi_1(k) + \phi_{-1}(k)) \right].
\end{equation}
In order to find the scattering eigenstates, we use the \emph{ansatz} for the photonic amplitude at sites $n\neq 0 $ as follows \cite{Zhou2008},
\begin{equation}
\phi_n(k) =
\begin{cases}
 e^{ikn} + r_ke^{-ikn}, & n<0\\
 t_ke^{ikn}, & n>0 
\end{cases}
\; .
\end{equation}
Using the continuity of the wave function,  $\phi_{0^{+}}(k) = \phi_{0^{-}}(k)$ the solution for the {transmittance amplitude} is:     
\begin{gather}
    \label{eq:transm}
    t_k = \frac{\Delta - \omega_k- \frac{g^2}{\omega^2_c}(\omega_c + \omega_k)}{\omega_k - \Delta - i \frac{g^2\omega^2_k/\omega^2_c}{\sqrt{(2\xi)^2-(\omega_k-\omega_c)^2}} +\frac{g^2}{\omega^2_c}(\omega_k + \omega_c)}.
\end{gather}
\begin{figure}[t]
    \centering
    \includegraphics[width = \columnwidth]{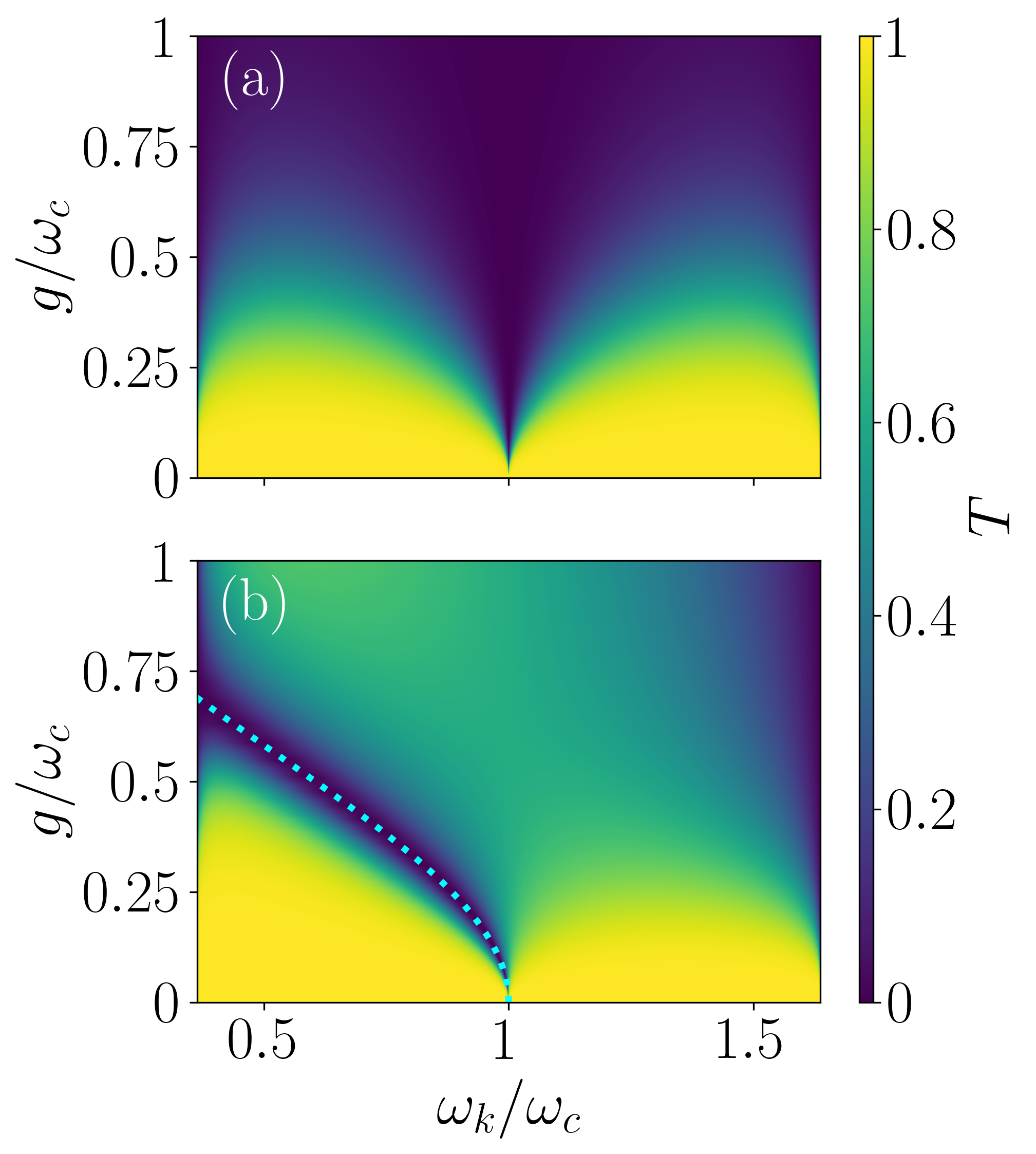}
    \caption{Transmittance spectra of a single photon within the rotating-wave approximation in both the truncated Coulomb gauge (a) and the truncated dipole gauge (b). The dipole gauge resonant frequency obtained in Eq. \eqref{eq:ZeroTRWA} is shown as a dotted cyan line. The parameters defining the system are the same as in Fig. \ref{fig:Spectra}.
    }
    \label{fig:Transm_RWA}
\end{figure}

An equivalent computation can be performed for the Coulomb gauge, assuming the same approximations mentioned in \ref{sec:wQED_H} to obtain a spin-boson model, this is, neglecting the $A^2$ term and considering the RWA.
By applying these transformations to Hamiltonian \eqref{eq:HCoulomb}, we obtain a model that also conserves the number of excitations. Further details about the resulting model can be found in Appendix \ref{app:weak}.
With this derived number-conserving description of our system in the Coulomb gauge, we can use the \emph{ansatz} \eqref{eq:SingleExcAntz} to obtain $t_k$. It yields the following expression (compare with Eq. \eqref{eq:transm})
\begin{equation}
\label{eq:transmcoul}
    t_k = \frac{\Delta- \omega_k}{\omega_k - \Delta - \frac{ig^2_C}{\sqrt{(2\xi)^2 - (\omega_k - \omega_c)^2}}},
\end{equation}
where $g_C = qA_0\braket{0\lvert p}{1}/m$ is the coupling strength in the Coulomb gauge, see App. \ref{app:weak} for more details.

Figure \ref{fig:Transm_RWA} compares both formulas \eqref{eq:transm} and \eqref{eq:transmcoul}, with {the transmittance} $T = \lvert t_k\rvert^2$, as a function of the coupling strength $g$ in the dipole gauge and the incoming photon frequency $\omega_k$.
In the truncated Coulomb gauge, it is a known result \cite{Zhou2008, Sanchez-Burillo2014} that the resonance frequency always occurs at the dipole transition, and the width of the transmittance minima increases as $g^2$ as plotted in Fig. \ref{fig:Transm_RWA} (a).

The equivalent computation in the truncated dipole gauge is presented in Figure \ref{fig:Transm_RWA} (b).
A notable feature in the dipole gauge is that the resonance moves to lower frequencies as $g$ increases. The resonant frequency can be obtained by imposing $t_k=0$ in Eq. \eqref{eq:transm}, leading to the formula:
\begin{equation}
    \label{eq:ZeroTRWA}   \omega^{\mathrm{RWA}}_{\mathrm{res}} = \omega_c \frac{\Delta\omega_c -g^2}{\omega^2_c + g^2}
    \; .
\end{equation}
This dependence is represented by the dotted cyan line in Figure \ref{fig:Transm_RWA} (b).
The resonant frequency shift is a consequence of the coupling-dependent term in the numerator of Eq. \eqref{eq:transm}, which arises from the couplings to the adjacent cavities in the real-space Hamiltonian \eqref{eq:HDipole-position}.
This frequency shift will be confirmed with the full model in the next section.
Another effect of the correct truncation is the modification of the width of the resonance, as given by the imaginary term in \eqref{eq:transm}.
In Figure \ref{fig:Transm_RWA} (b), we can observe a transmittance imbalance at both sides of the red-shifted minima. For a given coupling value, the transmittance is higher at frequencies below the resonance than at frequencies above.

To gain a deeper understanding of the resonance shift, we employ the resolvent operator method \cite{Cohen-AtPhInt-Ch3}. 
This approach allows us to identify the change in resonance as a Lamb shift.
The self-energy of our spin-boson model within the RWA can be expressed as:
\begin{equation}
    \label{eq:resolvent}
    \Sigma(E) = \sum_k \frac{\lvert g_k \rvert^2}{E-\omega_k} = \frac{g^2}{N\omega^2_c} \sum_k \frac{\omega^2_k}{E-\omega_k} \; .
\end{equation}
In the continuum limit, this summation can be written in terms of known integrals \cite{Economou2006, Lombardo2014}. 
Further details on this computation can be found in Appendix \ref{sec:app_SelfEn}.
After performing some manipulations, the self-energy of the system can be expressed as follows:
\begin{equation}
    \label{eq:SelfEn}
    \Sigma(E) = -\frac{g^2}{\omega^2_c}  (\omega_c + E) + i \frac{g^2}{\sqrt{(2\xi)^2 - (E-\omega_c)^2}} \frac{E^2}{\omega^2_c} \; .
\end{equation}
The real part of the self-energy represents the Lamb shift, causing the red-shift proportional to the square of the normalized coupling strength $g/\omega_c$, as given in Eq. \eqref{eq:transm}.
Similarly, the imaginary part of \eqref{eq:SelfEn} corresponds to half of the spontaneous emission rate, which can also be obtained by evaluating the spectral density $J_D(\omega)$ at the dipole transition ${\Delta}$, as shown in Eq. \eqref{eq:JD}.

The shift in the resonance frequency \eqref{eq:ZeroTRWA} is one of the primary findings of this study.
Although it has been calculated under the rotating-wave approximation, we will observe a similar effect in the full model.
Importantly, this shift is a measurable quantity that sheds light on the issues associated with truncating the matter in the Coulomb gauge.


%
%
%
%

\subsection{Beyond RWA}
\label{sec:R_USC}

Having established that the truncation within the Coulomb gauge cannot account for the Lamb shift and the resonant frequency dependence on the coupling, we now turn our attention to the full model without applying the RWA.
This allows us to study the transmission in a wide range of parameters.

To ensure the accuracy of our results and interpretations, we performed simulations using the matching technique by truncating to different numbers of levels in the dipole gauge. 
Additionally, we conducted MPS simulations, as presented in Appendix \ref{sec:app_CoulombT}, where it can be observed that truncating the dipole gauge to two levels yields the correct results. 
Therefore, in the main text, we will only describe this specific case.
Without the RWA, the number of excitations is no longer conserved. 
As a result, processes that involve different numbers of photons between the input and output fields become possible.
However, after conducting thorough numerical investigations, we have found that these processes have negligible magnitudes. 
Therefore, in this study, we can focus on the single photon transmission without significant loss of accuracy.

\begin{figure}[t]
    \centering
    \includegraphics[width = \linewidth]{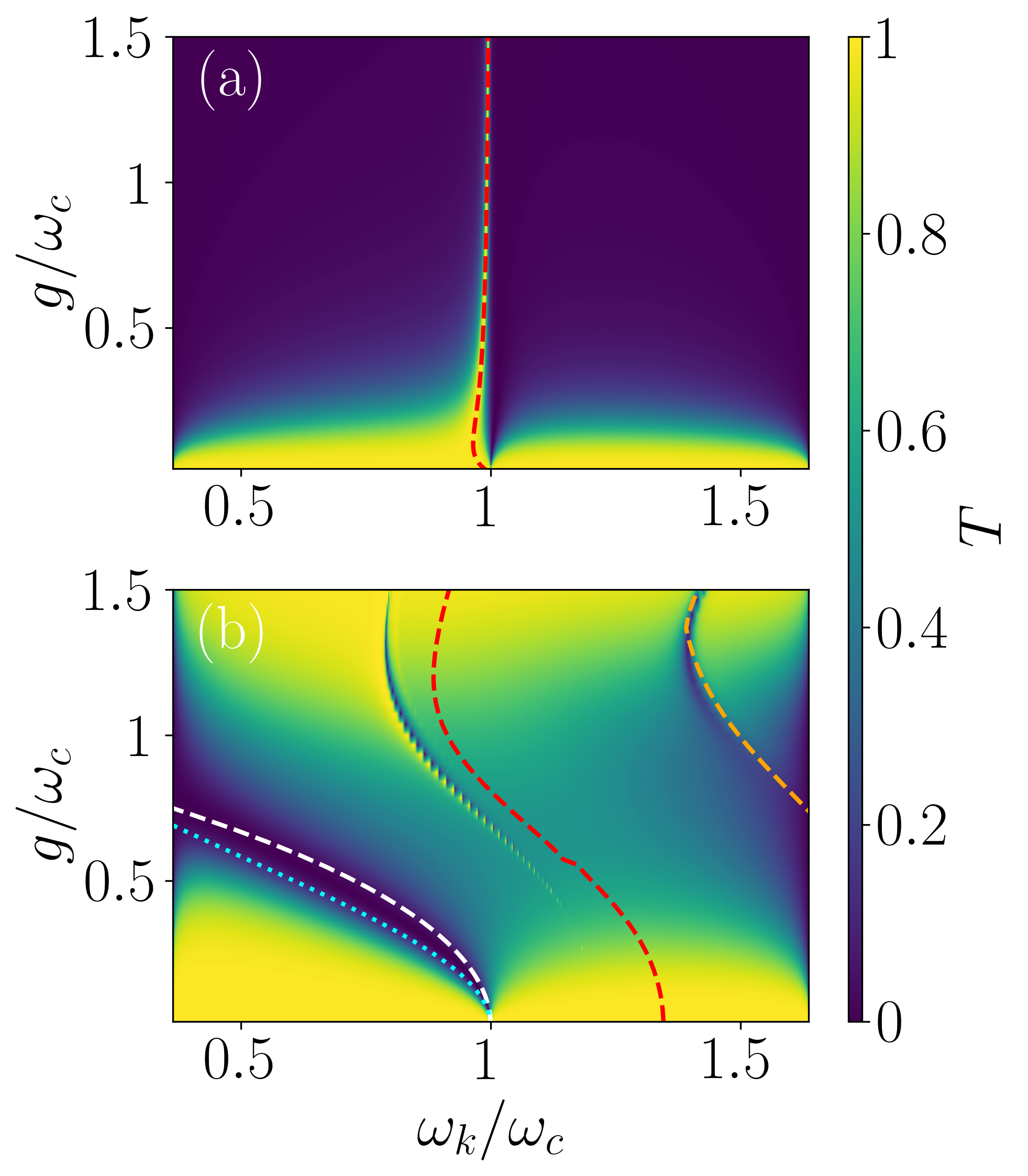}
    \caption{
   (a) Elastic transmittance spectra for a single-photon propagating in a coupled cavity array, including all the interaction terms of the truncated Coulomb gauge, as a function of the dipole gauge coupling strength and the incoming photon frequency. 
    The red dashed line indicates the scatterer transition relevant to the photon transport in this gauge.    
    (b) Equivalent spectra computed in the dipole gauge.
    %
   The red and orange dashed lines depict the transition energies of the scatterer in the dipole gauge that play a role in the transmission process.
    Additionally, we plot the resonance predicted in the rotating-wave approximation in a dotted cyan line, as given in equation \eqref{eq:ZeroTRWA}. 
    The resonance predicted utilising polaron techniques, as described in section \ref{sec:results}, is depicted with a dashed white line.}
    \label{fig:T_USC}
\end{figure}

The complete {elastic} transmittance spectra obtained in the Coulomb and dipole gauges are depicted in Figures \ref{fig:T_USC} (a) and (b), respectively.
In the Coulomb gauge spectra shown in Figure \ref{fig:T_USC} (a), the transmittance minima remains constant, much like in the number-conserving transmittance spectra in Figure \ref{fig:Transm_RWA} (a).
One notable feature arising from the counterrotating terms in the Coulomb gauge is a Fano resonance near the center of the band for all coupling strengths.

The corresponding transmittance spectra in the dipole gauge are plotted in Figure \ref{fig:T_USC} (b).
Several features can be observed in this plot.
First, let's discuss the resonance frequency shift, which already occurs within the RWA, as detailed above. 
To understand this shift in the full model, we can employ the polaron picture and the effective Hamiltonian $\mathcal{H}_P$ \eqref{eq:HPol}.
This transformed Hamiltonian is number-conserving, enabling us to compute the self-energy similarly to the previous section. In this case, the self-energy is given by:
\begin{align}
    \label{eq:SelfEnPol}
    \Sigma_P(E) = \sum_k \frac{4\Delta^2_r \lvert f_k\rvert^2}{(E-\omega_k - 2\Delta_r \sum_{l,p}f_l f_p )} \, .
\end{align}
The equation for the resonance is now given by [Cf. with eq. \eqref{eq:ZeroTRWA}]:
\begin{equation}
    \omega_k -\Delta_r + \Re \left( \Sigma_P(\omega_k)\right) = 0 \, .
\end{equation}
Here, $\Delta_r$ represents the renormalized transition frequency \eqref{Deltar}.
In Figure \ref{fig:T_USC} (b), the frequency shift is depicted as a white dashed line, matching the numerical results, and for comparison, we also plot the RWA result \eqref{eq:ZeroTRWA} with a cyan dotted line.
The shift is smaller than in the RWA case, which is due to the competition of two effects. 
On the one hand, we have the renormalized 
dipole frequency $\Delta_r$ from \eqref{Deltar}.
On the other hand, there is $\Re \left( \Sigma_P(\omega_k)\right)$, which tends to shift towards higher frequencies.

The main result of this paper is confirming the resonance shift in the full model, which provides a qualitatively measurable feature to test gauge issues related to the {truncation of dipole energies}.

\begin{figure}[t]
    \centering
    \includegraphics[width = \columnwidth]{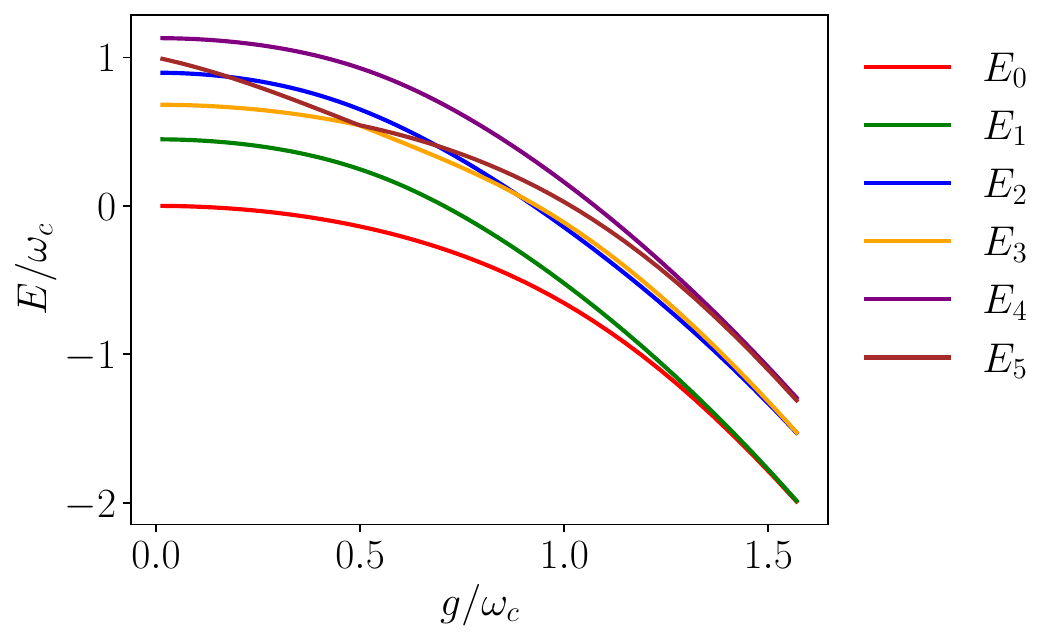}
    \caption{Lowest eigenenergies of Region (II),  which includes the dipole and the 5 central cavities, as introduced in \ref{sec:M_Matching}.}
    \label{fig:ScatEns}
\end{figure}

Furthermore, Fano resonances also appear in the spectra. 
The first resonance occurs at intermediate frequencies and spans most of the range of $g$ values.
Another resonance appears at larger coupling strengths and for the higher frequencies of the band.
These Fano resonances can be explained by considering the interaction between the flying photon and the dipole, which is allowed to access subspaces with a larger number of excitations due to the inclusion of counterrotating terms \cite{Sanchez-Burillo2014, Vrajitoarea2022}.
As a result, the flying photon can resonate with eigenstates of the scatterer (defined as region (II) in Sec. \ref{sec:M_Matching}) having different numbers of excitations.
In Figure \ref{fig:ScatEns}, we plot the eigenstates of region II as a function of the coupling strength. 
Two localized eigenstates with odd parity, corresponding to three and five excitations, are identified and denoted as $E_3$ and $E_5$, respectively.
We also plot the energy differences between these states and the ground state, i.e., $E_3 - E_0$ and $E_5 - E_0$, in Figure \ref{fig:T_USC} (b) as dashed lines in red and orange, respectively.
A similar analysis can be done in the Coulomb gauge, where a three-excitation eigenstate can be associated with the Fano resonance observed in that spectrum.
The agreement with the numerical results confirms our argument explaining the presence of the Fano resonances.
%

%
%

\begin{figure}[t]
    \centering
    \includegraphics[width = \columnwidth]{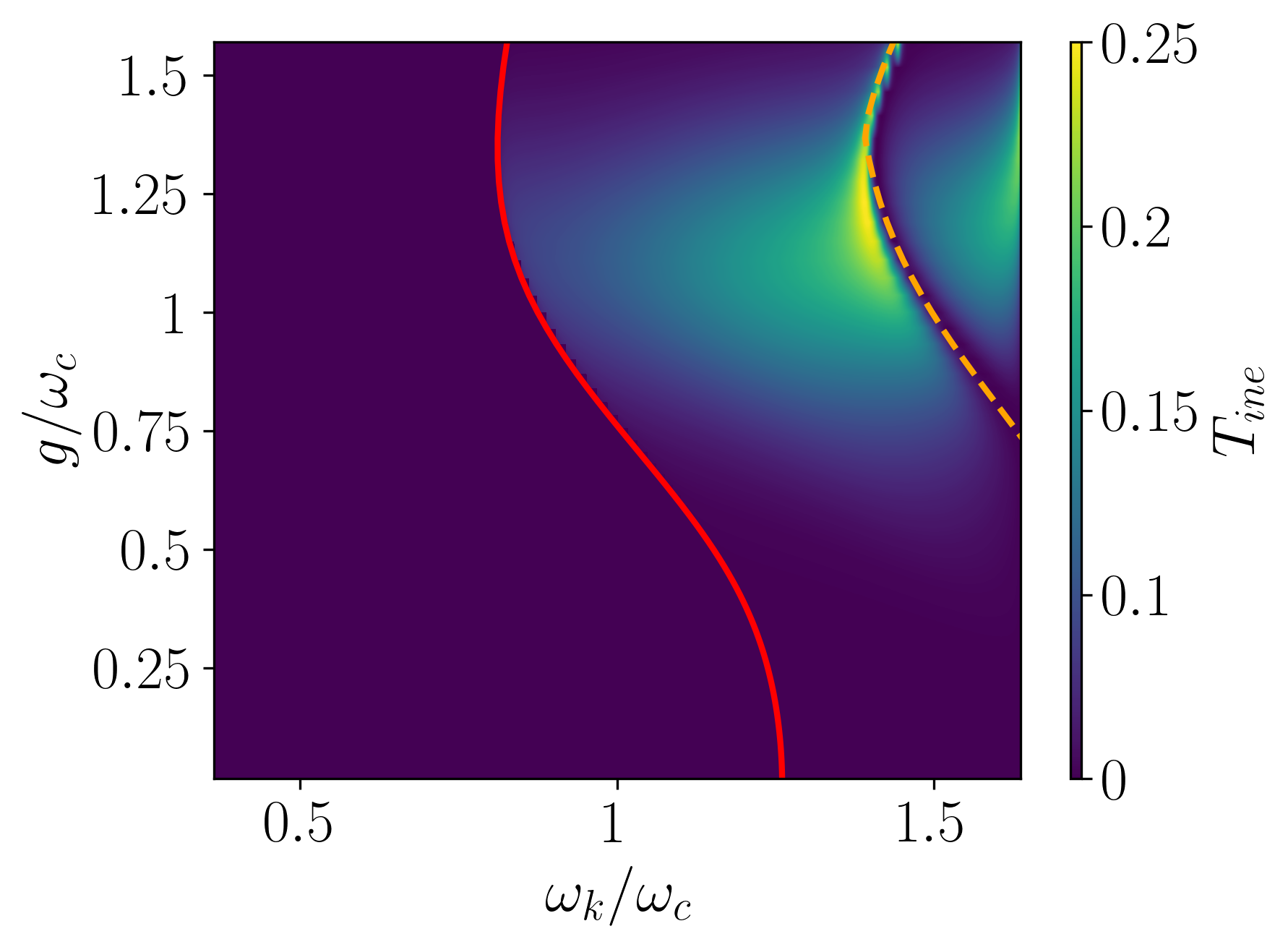}
    \caption{Inelastic transmittance 
    $(1-T-R)/2$ in the dipole gauge. The orange dashed line indicates the same Fano resonance as in \ref{fig:T_USC}. The red solid line indicates the minimum energy of the incoming photon required for inelastic scattering to occur, as given by Eq. \eqref{eq:omega_min}.}
    \label{fig:InelastT}
\end{figure}

Lastly, the full model exhibits inelastic scattering.
The presence of bound states originating from the ultrastrong coupling allows for scattering processes that leave the scatterer in an excited state.
%
%
In Figure \ref{fig:InelastT}, we present the inelastic transmittance spectra obtained from the dipole gauge computation, which reaches a maximum value of $0.25$, a fundamental bound as explained in \cite{Sanchez-Burillo2014}.

These inelastic processes correspond to Raman scattering, leaving the dressed dipole in an excited bound state.
To delineate the parameter range where inelastic transmission occurs, we use the energy conservation condition:
\begin{equation}
\label{eq:Econs}
\omega^{in}_k + E_0 = E_n +\omega^{out}_k,
\end{equation}

where $\omega^{out}_k \in [\omega_c - 2\xi, \omega_c + 2\xi]$, and $E_0$ and $E_n$ are the energies of the ground state and bound states for the dressed dipole.
The solid red line in Figure \ref{fig:InelastT} (c) represents the minimum energy of the incoming photon for which inelastic scattering is possible, given by Eq. \eqref{eq:omega_min}:
\begin{equation}
    \label{eq:omega_min}
    \omega^{min}_{ine} = E_2-E_0 + \omega_c - 2\xi.
\end{equation}
Our numerical computations verify this condition, providing the separation line beyond which inelastic transmission is possible.

\section{Implementation}
\label{sec:implementation}

\begin{figure}
    \centering
    \includegraphics[width = \linewidth]{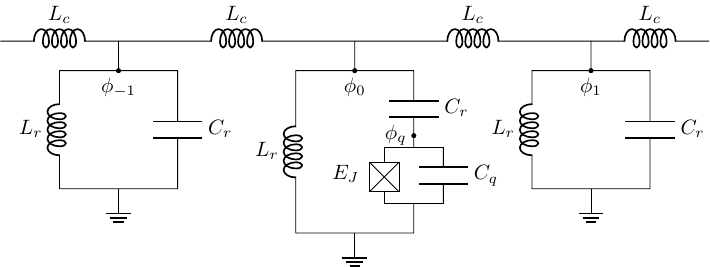}
    \caption{Cavity array implemented in a superconducting circuit QED platform. Individual LC oscillators are coupled inductively to their neighbours, while the transmon is directly connected to the capacitance of the central oscillator, labelled $0$.}
    \label{fig:CavityArraycQED}
\end{figure}

{Finally}, we propose a circuit QED architecture that provides an experimental platform to test the ideas developed in this paper.
%
%
Our proposed circuit, illustrated in Figure \ref{fig:CavityArraycQED}, consists of an array of $N$ LC circuits coupled inductively in series.
The central circuit contains a superconducting qubit (such as a transmon) capacitively coupled to the LC circuit, as shown in the same figure.

By considering the Kirchhoff equations of motion and selecting $\phi_0$ and $\phi_q$ as variables, we can express the current through the capacitor $C_r$ as $C_r (\ddot{\phi}_0 - \ddot{\phi}_q)$, which leads to a light-matter coupling via momentum, analogous to the minimal coupling in Eq. \eqref{Hc}. 
The Hamiltonian for this circuit is given by \cite{DevoretFluctuations, Garcia-Ripoll2022},
\begin{align}\label{eq:ClasIndFlux}
{H}_{Ch} =& \sum_n \left[\frac{Q^2_n}{2C_r} + \frac{\phi^2_n}{2L_{\Sigma}} + \frac{\phi_n \phi_{n-1}}{L_c} \right] \nonumber \\
&+ \frac{(Q_0+Q_q)^2}{2C_J} + E_J \cos(\phi_q ) \; ,
\end{align}
where $1/2L_{\Sigma} = 1/2L_r + 1/L_c$. The suffix $Ch$ indicates that this Hamiltonian is obtained in the charge gauge, resulting from the choice of dynamical variables $\phi_0$ and $\phi_q$. 
The presence of the minimal coupling-like feature in this circuit makes it analogous to the light-matter Hamiltonian in the Coulomb gauge \eqref{eq:HCoulomb}.

Similar to dipolar systems, a unitary transformation can also be applied in this case (cf. Eq. \eqref{U}) \cite{DeBernardis2018a}:
\begin{equation}
\label{Ucircuit}
    U = e^{i\phi_q Q_0/\hbar} \; .
\end{equation}
Physically, this transformation changes the dynamical variables (for the Kirchhoff equations) from $\{\phi_0, \phi_q \}$ to $\{ \delta \phi, \phi_q \}$, where $\delta \phi = \phi_0 - \phi_q$.

By utilizing the mode fluxes and charges, $\phi_k = \frac{1}{\sqrt{N}} \sum_n \phi_n e^{i k n}$, $Q_k = \frac{1}{\sqrt{N}} \sum_n Q_n e^{ikn}$, and $\alpha_k = \left(\frac{1}{2L_{\Sigma}} + \frac{\cos(k)}{L_c}\right)$, along with their quantization \cite{DevoretFluctuations, Garcia-Ripoll2022}:
\begin{align*}
    Q_k &= -i \sqrt{\frac{\hbar \omega_k C_r}{2}} (a^\dagger_k - a_k),\\
    \phi_k &= \sqrt{\frac{\hbar \omega_k}{4\alpha_k}}(a^\dagger_k + a_k) \; ,
\end{align*}
the transformed Hamiltonian $H_{Fl} = U H_{Ch} U^\dagger$ can be written as:
\begin{widetext}
\begin{align}
    \label{eq:cQED_Hamiltonian}
    H_{Fl} =
    \frac{Q^2_q}{2C_J} + E_J \cos\left(\frac{\phi_q - \phi_{ext}}{\Phi_0}\right)
    +
    \sum_k {\hbar \omega_k} a^\dagger_k a_k  - \hbar\sum_k g_k(a_k+ a^\dagger_k) \phi_q 
    + \frac{1}{N} \sum_k \alpha_k \lvert \phi_q \rvert^2 
     \; .
\end{align}
\end{widetext}
Here, the dispersion relation is $\omega_k = \omega_r + 2\xi_r\cos(k)$, where $\omega_r = (L_\Sigma C_r)^{-1/2}$ and $\xi_r = \omega_r L_\Sigma/L_c$.
The coupling constants are given by:
\begin{equation}
\label{gk_circuit}
\hbar g_k = \sqrt{\frac{\hbar}{2L_\Sigma \omega_cN}}\omega_k.
\end{equation}
It is worth noting that $g_k \sim \omega_k$ as shown in Eq. \eqref{eq:gk}.
Moreover, the bare matter Hamiltonian, defined by the first two terms in \eqref{eq:cQED_Hamiltonian}, is also modified by a term scaling with the square of the flux operator. 
Therefore, this circuit Hamiltonian is equivalent to \eqref{HD}.

%
%
%
%
\section{Conclusions}
\label{sec:Conclusions}

In this work, we have extended the study of gauge issues in light-matter coupled systems to the realm of waveguide QED. 
Gauge problems present significant technical challenges when studying waveguide QED, particularly in the ultrastrong coupling regime.
We have employed various theoretical methods, including numerical techniques (matching and MPS) and analytical approaches (polaron transformation), to describe the system dynamics and compare truncation in different gauges.
We argued that scattering, a natural quantity in waveguide QED, is ideal for testing different gauges, and it holds relevance from an experimental perspective.


Our investigations have confirmed that the transmittance spectrum exhibits both qualitative and quantitative differences when truncating in different gauges.
Numerical results have provided evidence that the dipole gauge is well-suited for truncation, allowing for accurate transmittance spectra over a wide parameter range.
On the other hand, the Coulomb gauge is found to be unsuitable for truncation.
Figure \ref{fig:T_USC} presents a clear visual representation of these significant differences.

The main features of correct transmittance spectra include the coupling-dependent resonant frequency shift, the emergence of Fano-like resonances, and the occurrence of non-elastic scattering.
These findings constitute important aspects of our study.
Furthermore, we have concluded the article by proposing an experimental implementation of this physics using circuit QED.

\section{Acknowledgments}

The authors acknowledge funding from the Spanish Government Grants PID2020-115221GB-C41/AEI/10.13039/501100011033 and TED2021-131447B-C21 funded by MCIN/AEI/10.13039/501100011033 and the EU ``NextGenerationEU''/PRTR, the Gobierno de Arag\'on (Grant E09-17R Q-MAD) and the CSIC Quantum Technologies Platform PTI-001.

\appendix

%
%

\section{Gauge Invariance at weak coupling}
\label{app:weak}

In this section, we demonstrate that in the weak coupling limit, both the Coulomb gauge and dipole gauge formulations can be represented as spin-boson models [see Eq. \eqref{HDtls}].
Although their spectral densities are not identical, under the Wigner-Weisskopf approximation, their spontaneous emission rates are equal, as indicated in Eq. \eqref{eq:LimitSpectrDens}.

By substituting the expression of the vector potential in the Coulomb gauge, as given in the main text equation \eqref{ACQ}, into the minimal coupling Hamiltonian \eqref{minimal}, we obtain a similar expression to that in the dipole gauge \eqref{HDsb}\begin{widetext}
\begin{equation}
\label{HCsb}
    H_C = H_{\rm m} + \sum_k \omega_k a_k^\dagger a_k 
    - \frac{q}{\sqrt{L}} \frac{{\bf p}}{m} \sum_k  \Big ( {\boldsymbol \lambda}_k  e^{i k z_0} a_k - {\rm h.c.} \Big ) +
     \frac{q^2}{2mL }   \Big( 
    \sum_k{\boldsymbol \lambda}_k  e^{i k z_0} a_k + {\rm h.c.}  \Big)^2 .
\end{equation}
\end{widetext}

To derive effective models in the weak coupling regime we consider the limit $\boldsymbol{\lambda}_k \to 0$ in both \eqref{HDsb} and \eqref{HCsb} retaining only the first-order terms in $\boldsymbol{\lambda}_k$.
Therefore neglecting the shift in $H_m$ for the dipole gauge and the ${\bf A}^2$-term in the Coulomb gauge.

After taking this limit, we project the effective Hamiltonians into the subspace spanned by the first two energy levels of the bare matter Hamiltonian $H_m$, denoted as ${\ket{0}, \ket{1} }$.
This procedure results in two spin-boson models, one for each gauge, which can be written in a form similar to the full model presented in the main text \eqref{HDtls}.

In this weak coupling limit, the waveguide modes $\omega_k$ and the two-level system transition frequencies are equal for both models.
%
%
%
The difference between the weak coupling Coulomb and dipole spin-boson models lies in their mode couplings.
In the dipole gauge, the expression for the full model couplings is given in \eqref{eq:gk_gen}.
For the approximated model in the dipole gauge, we have
\begin{equation}
    \label{eq:gkDip}
    g_{{\bf x},k} = \frac{q\omega_k}{\sqrt{L}}\langle 0 | {\bf x} | 1 \rangle \cdot {\boldsymbol \lambda}_k.
\end{equation}
The corresponding couplings in the Coulomb gauge can be written as
\begin{equation}
    \label{eq:gkCoul}
    g_{{\bf p},k} = \frac{q}{m\sqrt{L}}\langle 0 | {\bf p} | 1 \rangle \cdot {\boldsymbol \lambda}_k 
\end{equation}

We can now compute the spectral densities of these two models using the definition given by Eq. \eqref{eq:SpectrDensDef} in the main text, obtaining
\begin{align}
    J_D(\omega) &= \frac{q^2\omega^2}{L}\frac{\big|\langle 0 | {\bf x} | 1 \rangle \cdot {\boldsymbol \lambda_k}\big |^2}{\sqrt{4\xi^2 - (\omega - \omega_c)^2}},\\
    J_C(\omega) &= \frac{q^2}{m^2 L} \frac{\big|\langle 0 | {\bf p} | 1 \rangle \cdot {\boldsymbol \lambda_k}\big |^2}{\sqrt{4\xi^2 - (\omega - \omega_c)^2}}.
\end{align}
    
Utilizing the general relation
\begin{equation}
    \label{eq:Relop_px}
    \langle n |{\bf p} |l\rangle = i m \Delta_{nl}\langle n |{\bf x} |l\rangle,
\end{equation}
where $\Delta_{nl}$ is the transition frequency between states $\ket{n}$ and $\ket{l}$,
we can derive the following expression for the spectral density in the Coulomb gauge
\begin{equation}
    \label{eq:JCx}
    J_C(\omega) = \frac{q^2 \Delta^2}{L}\frac{\big|\langle 0 | {\bf x} | 1 \rangle \cdot {\boldsymbol \lambda_k}\big |^2}{\sqrt{4\xi^2 - (\omega - \omega_c)^2}}.
\end{equation}
This proves that in the weak coupling limit, the spontaneous emission rate $J(\Delta)$ is gauge invariant
\begin{equation}
    \label{eq:RelSpectralApp}
    \lim_{{\boldsymbol \lambda_k }\to 0 } J_C(\Delta) = \lim_{{\boldsymbol \lambda_k }\to 0 }J_D(\Delta).
\end{equation}

%
%
%
%
\section{Analytical computation of the Self-Energy of the coupled dipole}
\label{sec:app_SelfEn}
In this appendix, we provide the derivation of the exact self-energy of the dipole excited level in the RWA \eqref{eq:SelfEn}. 
We begin the derivation from the expression  \eqref{eq:resolvent} and assume that the energy $E$ lies within the band $E \in [\omega_c -2\xi, \omega_c+2\xi]$. 
However, the computation can be easily extended to energies outside the band.

As mentioned in the main text, expanding the expression of the dispersion relation in Eq. \eqref{eq:resolvent} yields:
\begin{align}
    \label{eq:SelfEnApp1}
    \Sigma(E) &= \frac{g^2}{N\omega^2_c} \sum_k \frac{\omega^2_k}{E-\omega_k} = \nonumber\\
    &=\frac{g^2}{N\omega^2_c} \sum_k \Bigg[\frac{\omega^2_c}{E-\omega_k} + \frac{2\xi\omega_c(e^{ik} + e^{-ik})}{E-\omega_k} \nonumber \\
    &+\frac{\xi^2\left(e^{ik} + e^{-ik} \right)^2}{E-\omega_k} \Bigg].
\end{align}

We can convert the summations over modes into integrals by taking the continuum limit. ($N \to \infty$).
\begin{align}
    \label{eq:SelfEnApp2}
    \Sigma(E) &= \frac{g^2}{2\pi\omega^2_c}\int^{\pi}_{-\pi} \Bigg[ \frac{(\omega^2_c +2\xi^2)dk}{E-\omega_k} +\frac{2\xi\omega_c (e^{ik} + e^{-ik})dk}{E-\omega_k}\nonumber\\
    &+ \int^{\pi}_{-\pi} \frac{\xi^2(e^{2ik} + e^{-2ik})dk}{E-\omega_k} \Bigg].
\end{align}
The entire computation in \eqref{eq:SelfEnApp2} relies one valuating integrals of the form:
\begin{align}
    I(E,n) =
    &\int^{\pi}_{-\pi} \frac{e^{ink} dk}{E-[\omega_c -\xi(e^{ik}+ e^{-ik})]}.
\end{align}
By performing the variable change $z = e^{ik}$, transforms into a closed integral around a circle of radius $\lvert z \rvert = 1$ in the complex plane
\begin{equation}
    \label{eq:GenericIEn}
    I(E,n) =\frac{1}{\xi} \oint_{\lvert z \rvert = 1}\frac{z^n dz}{z^2 + 2az + 1}
\end{equation}
where we defined $a = (E-\omega_c)/(2\xi)$.
Since the integral is around a closed circle, we have $I(E,n) = I(E,-n)$.

Integral \eqref{eq:GenericIEn} can be solved using the Residue Theorem.
If the energy lies within the band ($\lvert a\rvert <1$), the corresponding poles are $z_\pm = - a \pm i \sqrt{1-a^2}$. These poles correspond to the limits $\lim_{\eta \to 0} E\pm i \eta$. 
\begin{equation}
    \label{eq:SolIntGen}
    \lim_{\eta \to 0} I(E\pm i\eta, n) = \mp 2\pi i \frac{\left(-a \pm i\sqrt{1-a^2} \right)^{\lvert n \rvert}}{\sqrt{4\xi^2-(E-\omega_c)^2}}.
\end{equation}

Writing the self-energy \eqref{eq:SelfEnApp2} in terms of the integrals \eqref{eq:GenericIEn} gives:
\begin{align}
    \label{eq:SelfEnsum}
    \Sigma(E) = i\frac{g^2}{2\pi\omega^2_c}\Big[(\omega^2_c +2\xi^2)I(E,0) +\nonumber\\
    4\xi \omega_c I(E,1)+
    2\xi^2 I(E,2)
    \Big].
\end{align}

From where, after introducing the solution found in \eqref{eq:SolIntGen}, we can obtain the self-energy as in Eq. \ref{eq:SelfEn} of the main text. 
The real part of the self-energy gives the Lamb shift of the resonant frequency, whereas the imaginary part provides half its spectral width.

\begin{subequations}
\begin{equation}
        \text{Re} (\Sigma(E)) = -\frac{g^2}{\omega^2_c} (\omega_c + E).
\end{equation}
\begin{equation}
        \text{Im} (\Sigma(E)) = \frac{g^2}{\omega^2_c} \frac{E^2}{\sqrt{4\xi^2 - (E-\omega_c)^2}} = \frac{J(E)}{2}.
\end{equation}
\end{subequations}
%
%
%
%

\section{Transmission in the Coulomb gauge}

\label{sec:app_CoulombT}

In this appendix, we demonstrate the effect of truncation in both gauges by explicitly computing different transmission spectra while considering various numbers of dipole levels $N_d$.
As mentioned in the main text, applying the two-level approximation in the dipole gauge can accurately approximate the full model, whereas doing so in the Coulomb gauge does not yield the expected results.
However, we were able to perform some calculations in the Coulomb gauge for $N_d > 2$ at the intermediate range of the USC regime. 
\begin{figure}
    \centering
    \includegraphics[width = \columnwidth]{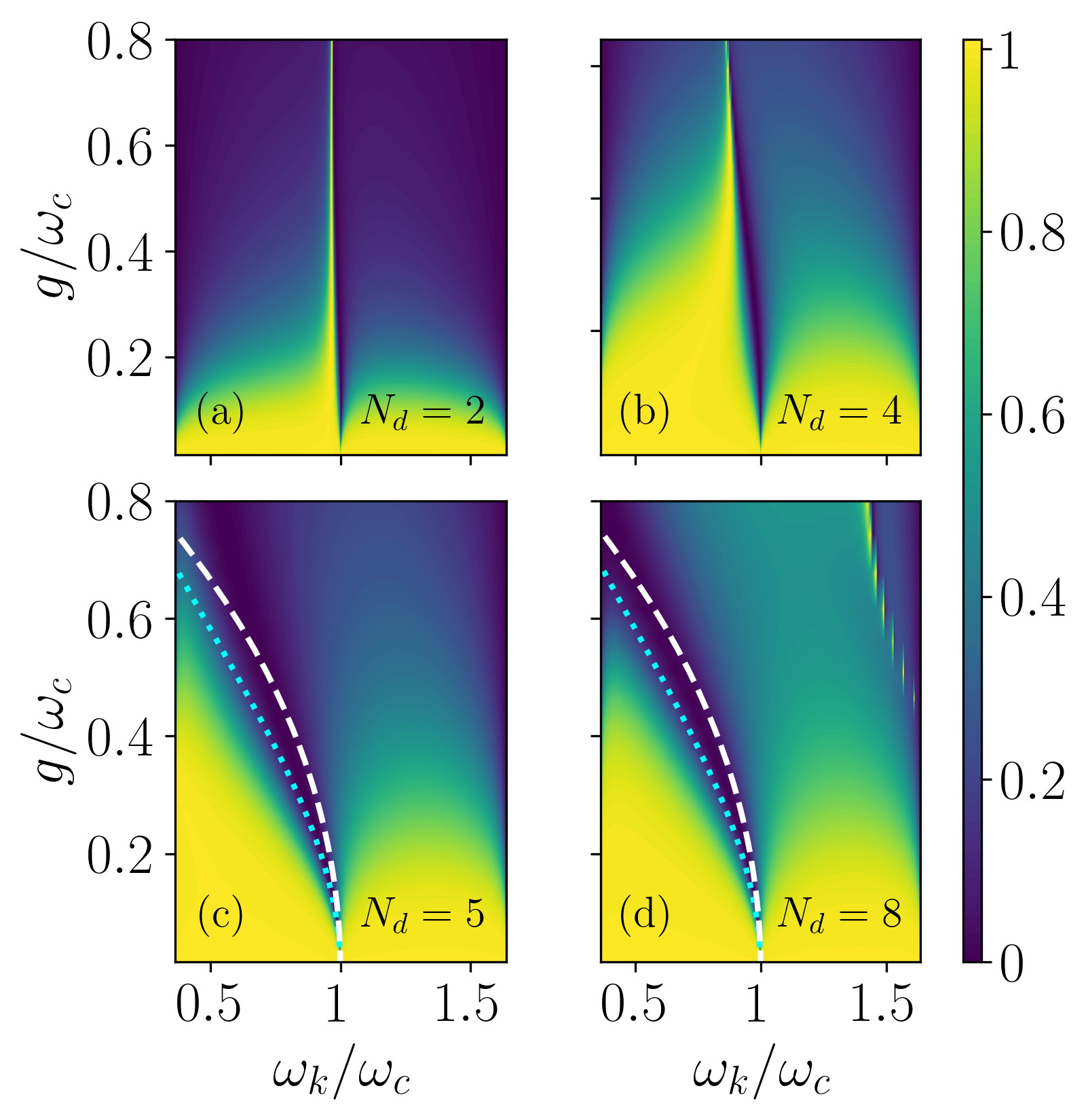}
    \caption{Transmittance spectra in the Coulomb gauge for a different number of dipole levels $N_d$. (a) and (b) correspond to the spectra obtained using the two-level approximation and for four levels, respectively. For a higher number of dipole levels, such as in (c) $N_d = 5$ and (d) $N_d = 8$, we also plot the RWA resonance as a dotted cyan line [Cf. Eq. \eqref{eq:ZeroTRWA}] and the USC resonance line predicted with the polaron transformation in the dipole gauge.}
    \label{fig:CoulombT}
\end{figure}

In Fig. \ref{fig:CoulombT}, we plot the transmittance spectra in the Coulomb gauge, including all coupling terms, for different numbers of dipole levels.
Fig. \ref{fig:CoulombT} (a) shows the spectra obtained using the two-level approximation in the Coulomb gauge, where the resonance is fixed at $\omega_k = \Delta$ and an additional Fano resonance is observed for all couplings.
In Fig. \ref{fig:CoulombT} (b), we increase the number of dipole levels, resulting in a slight red-shift of the main resonance and an increase in the transmittance at lower frequencies around the main resonance while maintaining the Fano resonance.
As the number of dipole levels increases further in Fig. \ref{fig:CoulombT} (c) and (d), we observe a strong red-shift in the transmittance minima and an imbalance of the transmittance on both sides of the main resonance. 
These effects are consistent with what we found and explained in the main text for the dipole gauge.

In both Fig. \ref{fig:CoulombT} (c) and (d), we also plot the minima of transmittance obtained in the dipole gauge, as shown in the main text in Fig. \ref{fig:T_USC}. 
The blue dotted line represents the analytical solution in the RWA, as given in Eq. \eqref{eq:ZeroTRWA}, while the white dashed line gives the solution in the effective polaron picture.
It is evident that capturing the main features obtained in the dipole gauge with the two-level approximation in the Coulomb gauge requires a significantly higher number of dipole levels. 
While the main resonance seems to be well described, the Fano resonances do not coincide yet. 
The Fano resonances directly come from the eigenenergies of the scatterer, which are well approximated in the dipole gauge, as introduced in Sec. \ref{sec:wQED_Trunc}.

However, obtaining an accurate two-level approximation in the Coulomb gauge from the truncated dipole model is possible, as observed in the single cavity case \cite{DiStefano2019}.

Applying a truncated version of the Power-Zineau-Wooley transformation \eqref{U}, $\mathcal{U} = \exp(ig/\omega_c(a_0+a^\dagger_0)\sigma_x)$, we can recover a Coulomb gauge description from $\mathcal{H}_C = \mathcal{U}\mathcal{H}_D\mathcal{U}^\dagger$ [Cf. Eqs. \eqref{eq:H0_RWA}, \eqref{eq:H1_RWA}]
\begin{align}
    \label{eq:TruncC}
    \mathcal{H}_C &=  \sum_n \omega_c a^\dagger_n a_n 
    +\xi\sum_n (a^\dagger_{n+1} a_{n}+a^\dagger_n a_{n+1}) \\
    &+\frac{\Delta^\prime}{2} \left[ \cos\left(\frac{2g}{\omega_c}(a_0 + a^\dagger_0)\right) +  \sin\left(\frac{2g}{\omega_c}(a_0 + a^\dagger_0)\right)\right].\nonumber
\end{align}
    
%
%
%
%

\section{Continuous limit of the system}
\label{sec:app_ContLim}

Cavity array systems can be experimentally implemented in various platforms, such as photonic crystals and superconducting systems \cite{Liu2017, Vrajitoarea2022}, as discussed in Sec. \ref{sec:implementation}.
Moreover, these types of models can also be seen as discretizations of general waveguide models.

In this appendix, we derive the continuous real-space description of our system in the dipole gauge.
This model differs from other standard continuous waveguide QED models \cite{Shen2009} due to the couplings to adjacent cavities.

To begin, we split the momentum space into left and right propagating momenta as follows:
\begin{equation}
    \label{eq:PhContK}
    H_{ph} =  \sum_{j = \{L,R\}} \int{dk_j} \omega({k_j}) a^\dagger(k_j)a(k_j) 
\end{equation}
where $[a(k_i), a^\dagger(k^\prime_j)] = \delta (k_j - k^\prime_i)\delta_{ij}$.

Next, we introduce creation and annihilation operators for right or left propagating photons at position $x$ as the Fourier transform of their momentum counterparts
\begin{gather}
    a(k_R) := \int a_R(r) e^{-ik_R r} dr,\\
    a(k_L) := \int a_{L}(r) e^{-ik_L r} dr.
\end{gather}

By linearizing the dispersion relation around a probe wavevector $k_0$, 
\begin{gather}
    \omega_L(k) \simeq \omega_L(k_0) - v_g(k_0) k_L,\\
    \omega_R(k) \simeq \omega_R(k_0) + v_g(k_0) k_R,
\end{gather}
with $k_L = k - k_0$ and $k_R = k + k_0$, we can find an expression for Hamiltonian \eqref{eq:PhContK} in real space as
\begin{align}
    H_{ph} &=   \int dr  a^\dagger_R(r) \left( \omega_R(k_0) -i v_g\frac{\partial}{\partial r}\right) a_R(r) \nonumber \\
    &+\int dx  a^\dagger_L(r) \left( \omega_L(k_0) -i v_g\frac{\partial}{\partial r}\right) a_L(r).
\end{align}
where we have used the relation $k_R e^{ik_R r} = -i\partial_r e^{ik_R r}$.

After applying the same linearization to the interaction term the full Hamiltonian reads
\begin{align}
    H_D &= \sum_{j}\int dr  a^\dagger_j(r) \left( \omega_j(k_0) -i v_g\frac{\partial}{\partial r}\right) a_j(r)
    + H_{m} \nonumber \\ 
&+i{q {A}_0} x\sum_{j} \int dk_j \left( \omega_j(k_0) -i v_g\frac{\partial}{\partial r}\right) a_j(r) \nonumber\\
&-i{q {A}_0} x\sum_{j} \int dk_j \left( \omega_j(k_0) -i v_g\frac{\partial}{\partial r}\right) a^\dagger_j(r).
\end{align}

\bibliography{bibDipoleGauge.bib}
\end{document}